\documentclass[a4paper, traditabstract, longauth]{aa} 

\def\setsymbol#1#2{\expandafter\def\csname #1\endcsname{#2}}
\def\getsymbol#1{\csname #1\endcsname}

\def\Planck{\textit{Planck}}



\newbox\tablebox    \newdimen\tablewidth
\def\leaderfil{\leaders\hbox to 5pt{\hss.\hss}\hfil}
%
%

\def\endPlancktablewide{\tablewidth=\textwidth 
    $$\hss\copy\tablebox\hss$$
    \vskip-\lastskip\vskip -2pt}
\def\tablenote#1 #2\par{\begingroup \parindent=0.8em
    \abovedisplayshortskip=0pt\belowdisplayshortskip=0pt
    \noindent
    $$\hss\vbox{\hsize\tablewidth \hangindent=\parindent \hangafter=1 \noindent
    \hbox to \parindent{$^#1$\hss}\strut#2\strut\par}\hss$$
    \endgroup}
\def\doubleline{\vskip 3pt\hrule \vskip 1.5pt \hrule \vskip 5pt}

%
\def\L2{\ifmmode L_2\else $L_2$\fi}

\def\DeltaT{\ifmmode \Delta T\else $\Delta T$\fi}
\def\deltat{\ifmmode \Delta t\else $\Delta t$\fi}
\def\fknee{\ifmmode f_{\rm knee}\else $f_{\rm knee}$\fi}
\def\Fmax{\ifmmode F_{\rm max}\else $F_{\rm max}$\fi}
\def\solar{\ifmmode{\rm M}_{\mathord\odot}\else${\rm M}_{\mathord\odot}$\fi}
\def\Msolar{\ifmmode{\rm M}_{\mathord\odot}\else${\rm M}_{\mathord\odot}$\fi}
\def\Lsolar{\ifmmode{\rm L}_{\mathord\odot}\else${\rm L}_{\mathord\odot}$\fi}

\def\inv{\ifmmode^{-1}\else$^{-1}$\fi}
\def\mo{\ifmmode^{-1}\else$^{-1}$\fi}
\def\sup#1{\ifmmode ^{\rm #1}\else $^{\rm #1}$\fi}
\def\expo#1{\ifmmode \times 10^{#1}\else $\times 10^{#1}$\fi}
\def\,{\thinspace}
\def\lsim{\mathrel{\raise .4ex\hbox{\rlap{$<$}\lower 1.2ex\hbox{$\sim$}}}}
\def\gsim{\mathrel{\raise .4ex\hbox{\rlap{$>$}\lower 1.2ex\hbox{$\sim$}}}}

\def\simprop{\mathrel{\raise .4ex\hbox{\rlap{$\propto$}\lower 1.2ex\hbox{$\sim$}}}}
\def\deg{\ifmmode^\circ\else$^\circ$\fi}
\def\pdeg{\ifmmode $\setbox0=\hbox{$^{\circ}$}\rlap{\hskip.11\wd0 .}$^{\circ}
          \else \setbox0=\hbox{$^{\circ}$}\rlap{\hskip.11\wd0 .}$^{\circ}$\fi}
\def\arcs{\ifmmode {^{\scriptstyle\prime\prime}}
          \else $^{\scriptstyle\prime\prime}$\fi}
\def\arcm{\ifmmode {^{\scriptstyle\prime}}
          \else $^{\scriptstyle\prime}$\fi}
\newdimen\sa  \newdimen\sb
\def\parcs{\sa=.07em \sb=.03em
     \ifmmode \hbox{\rlap{.}}^{\scriptstyle\prime\kern -\sb\prime}\hbox{\kern -\sa}
     \else \rlap{.}$^{\scriptstyle\prime\kern -\sb\prime}$\kern -\sa\fi}
\def\parcm{\sa=.08em \sb=.03em
     \ifmmode \hbox{\rlap{.}\kern\sa}^{\scriptstyle\prime}\hbox{\kern-\sb}
     \else \rlap{.}\kern\sa$^{\scriptstyle\prime}$\kern-\sb\fi}
\def\ra[#1 #2 #3.#4]{#1\sup{h}#2\sup{m}#3\sup{s}\llap.#4}
\def\dec[#1 #2 #3.#4]{#1\deg#2\arcm#3\arcs\llap.#4}
\def\deco[#1 #2 #3]{#1\deg#2\arcm#3\arcs}
\def\rra[#1 #2]{#1\sup{h}#2\sup{m}}

\def\dots{\relax\ifmmode \ldots\else $\ldots$\fi}
%
%
\def\WHzsr{\ifmmode $W\,Hz\mo\,sr\mo$\else W\,Hz\mo\,sr\mo\fi}
\def\mHz{\ifmmode $\,mHz$\else \,mHz\fi}
\def\GHz{\ifmmode $\,GHz$\else \,GHz\fi}
\def\mKs{\ifmmode $\,mK\,s$^{1/2}\else \,mK\,s$^{1/2}$\fi}
\def\muKs{\ifmmode \,\mu$K\,s$^{1/2}\else \,$\mu$K\,s$^{1/2}$\fi}
\def\muKRJs{\ifmmode \,\mu$K$_{\rm RJ}$\,s$^{1/2}\else \,$\mu$K$_{\rm RJ}$\,s$^{1/2}$\fi}
\def\muKHz{\ifmmode \,\mu$K\,Hz$^{-1/2}\else \,$\mu$K\,Hz$^{-1/2}$\fi}
\def\MJysr{\ifmmode \,$MJy\,sr\mo$\else \,MJy\,sr\mo\fi}
\def\MJysrmK{\ifmmode \,$MJy\,sr\mo$\,mK$_{\rm CMB}\mo\else \,MJy\,sr\mo\,mK$_{\rm CMB}\mo$\fi}
\def\microns{\ifmmode \,\mu$m$\else \,$\mu$m\fi}

\def\muK{\ifmmode \,\mu$K$\else \,$\mu$\hbox{K}\fi}
\def\microK{\ifmmode \,\mu$K$\else \,$\mu$\hbox{K}\fi}
\def\muW{\ifmmode \,\mu$W$\else \,$\mu$\hbox{W}\fi}
\def\kms{\ifmmode $\,km\,s$^{-1}\else \,km\,s$^{-1}$\fi}
\def\kmsMpc{\ifmmode $\,\kms\,Mpc\mo$\else \,\kms\,Mpc\mo\fi}
%
%


\setsymbol{LFI:center:frequency:70GHz:units}{70.3\,GHz}
\setsymbol{LFI:center:frequency:44GHz:units}{44.1\,GHz}
\setsymbol{LFI:center:frequency:30GHz:units}{28.5\,GHz}

\setsymbol{LFI:center:frequency:70GHz}{70.3}
\setsymbol{LFI:center:frequency:44GHz}{44.1}
\setsymbol{LFI:center:frequency:30GHz}{28.5}

\setsymbol{LFI:center:frequency:LFI18:Rad:M:units}{71.7\GHz}
\setsymbol{LFI:center:frequency:LFI19:Rad:M:units}{67.5\GHz}
\setsymbol{LFI:center:frequency:LFI20:Rad:M:units}{69.2\GHz}
\setsymbol{LFI:center:frequency:LFI21:Rad:M:units}{70.4\GHz}
\setsymbol{LFI:center:frequency:LFI22:Rad:M:units}{71.5\GHz}
\setsymbol{LFI:center:frequency:LFI23:Rad:M:units}{70.8\GHz}
\setsymbol{LFI:center:frequency:LFI24:Rad:M:units}{44.4\GHz}
\setsymbol{LFI:center:frequency:LFI25:Rad:M:units}{44.0\GHz}
\setsymbol{LFI:center:frequency:LFI26:Rad:M:units}{43.9\GHz}
\setsymbol{LFI:center:frequency:LFI27:Rad:M:units}{28.3\GHz}
\setsymbol{LFI:center:frequency:LFI28:Rad:M:units}{28.8\GHz}
\setsymbol{LFI:center:frequency:LFI18:Rad:S:units}{70.1\GHz}
\setsymbol{LFI:center:frequency:LFI19:Rad:S:units}{69.6\GHz}
\setsymbol{LFI:center:frequency:LFI20:Rad:S:units}{69.5\GHz}
\setsymbol{LFI:center:frequency:LFI21:Rad:S:units}{69.5\GHz}
\setsymbol{LFI:center:frequency:LFI22:Rad:S:units}{72.8\GHz}
\setsymbol{LFI:center:frequency:LFI23:Rad:S:units}{71.3\GHz}
\setsymbol{LFI:center:frequency:LFI24:Rad:S:units}{44.1\GHz}
\setsymbol{LFI:center:frequency:LFI25:Rad:S:units}{44.1\GHz}
\setsymbol{LFI:center:frequency:LFI26:Rad:S:units}{44.1\GHz}
\setsymbol{LFI:center:frequency:LFI27:Rad:S:units}{28.5\GHz}
\setsymbol{LFI:center:frequency:LFI28:Rad:S:units}{28.2\GHz}

\setsymbol{LFI:center:frequency:LFI18:Rad:M}{71.7}
\setsymbol{LFI:center:frequency:LFI19:Rad:M}{67.5}
\setsymbol{LFI:center:frequency:LFI20:Rad:M}{69.2}
\setsymbol{LFI:center:frequency:LFI21:Rad:M}{70.4}
\setsymbol{LFI:center:frequency:LFI22:Rad:M}{71.5}
\setsymbol{LFI:center:frequency:LFI23:Rad:M}{70.8}
\setsymbol{LFI:center:frequency:LFI24:Rad:M}{44.4}
\setsymbol{LFI:center:frequency:LFI25:Rad:M}{44.0}
\setsymbol{LFI:center:frequency:LFI26:Rad:M}{43.9}
\setsymbol{LFI:center:frequency:LFI27:Rad:M}{28.3}
\setsymbol{LFI:center:frequency:LFI28:Rad:M}{28.8}
\setsymbol{LFI:center:frequency:LFI18:Rad:S}{70.1}
\setsymbol{LFI:center:frequency:LFI19:Rad:S}{69.6}
\setsymbol{LFI:center:frequency:LFI20:Rad:S}{69.5}
\setsymbol{LFI:center:frequency:LFI21:Rad:S}{69.5}
\setsymbol{LFI:center:frequency:LFI22:Rad:S}{72.8}
\setsymbol{LFI:center:frequency:LFI23:Rad:S}{71.3}
\setsymbol{LFI:center:frequency:LFI24:Rad:S}{44.1}
\setsymbol{LFI:center:frequency:LFI25:Rad:S}{44.1}
\setsymbol{LFI:center:frequency:LFI26:Rad:S}{44.1}
\setsymbol{LFI:center:frequency:LFI27:Rad:S}{28.5}
\setsymbol{LFI:center:frequency:LFI28:Rad:S}{28.2}


\setsymbol{LFI:white:noise:sensitivity:70GHz:units}{134.7\muKs}
\setsymbol{LFI:white:noise:sensitivity:44GHz:units}{164.7\muKs}
\setsymbol{LFI:white:noise:sensitivity:30GHz:units}{143.4\muKs}

\setsymbol{LFI:white:noise:sensitivity:70GHz}{134.7}
\setsymbol{LFI:white:noise:sensitivity:44GHz}{164.7}
\setsymbol{LFI:white:noise:sensitivity:30GHz}{143.4}


\setsymbol{LFI:white:noise:sensitivity:LFI18:Rad:M:units}{512.0\muKs}
\setsymbol{LFI:white:noise:sensitivity:LFI19:Rad:M:units}{581.4\muKs}
\setsymbol{LFI:white:noise:sensitivity:LFI20:Rad:M:units}{590.8\muKs}
\setsymbol{LFI:white:noise:sensitivity:LFI21:Rad:M:units}{455.2\muKs}
\setsymbol{LFI:white:noise:sensitivity:LFI22:Rad:M:units}{492.0\muKs}
\setsymbol{LFI:white:noise:sensitivity:LFI23:Rad:M:units}{507.7\muKs}
\setsymbol{LFI:white:noise:sensitivity:LFI24:Rad:M:units}{462.2\muKs}
\setsymbol{LFI:white:noise:sensitivity:LFI25:Rad:M:units}{413.6\muKs}
\setsymbol{LFI:white:noise:sensitivity:LFI26:Rad:M:units}{478.6\muKs}
\setsymbol{LFI:white:noise:sensitivity:LFI27:Rad:M:units}{277.7\muKs}
\setsymbol{LFI:white:noise:sensitivity:LFI28:Rad:M:units}{312.3\muKs}
\setsymbol{LFI:white:noise:sensitivity:LFI18:Rad:S:units}{465.7\muKs}
\setsymbol{LFI:white:noise:sensitivity:LFI19:Rad:S:units}{555.6\muKs}
\setsymbol{LFI:white:noise:sensitivity:LFI20:Rad:S:units}{623.2\muKs}
\setsymbol{LFI:white:noise:sensitivity:LFI21:Rad:S:units}{564.1\muKs}
\setsymbol{LFI:white:noise:sensitivity:LFI22:Rad:S:units}{534.4\muKs}
\setsymbol{LFI:white:noise:sensitivity:LFI23:Rad:S:units}{542.4\muKs}
\setsymbol{LFI:white:noise:sensitivity:LFI24:Rad:S:units}{399.2\muKs}
\setsymbol{LFI:white:noise:sensitivity:LFI25:Rad:S:units}{392.6\muKs}
\setsymbol{LFI:white:noise:sensitivity:LFI26:Rad:S:units}{418.6\muKs}
\setsymbol{LFI:white:noise:sensitivity:LFI27:Rad:S:units}{302.9\muKs}
\setsymbol{LFI:white:noise:sensitivity:LFI28:Rad:S:units}{285.3\muKs}

\setsymbol{LFI:white:noise:sensitivity:LFI18:Rad:M}{512.0}
\setsymbol{LFI:white:noise:sensitivity:LFI19:Rad:M}{581.4}
\setsymbol{LFI:white:noise:sensitivity:LFI20:Rad:M}{590.8}
\setsymbol{LFI:white:noise:sensitivity:LFI21:Rad:M}{455.2}
\setsymbol{LFI:white:noise:sensitivity:LFI22:Rad:M}{492.0}
\setsymbol{LFI:white:noise:sensitivity:LFI23:Rad:M}{507.7}
\setsymbol{LFI:white:noise:sensitivity:LFI24:Rad:M}{462.2}
\setsymbol{LFI:white:noise:sensitivity:LFI25:Rad:M}{413.6}
\setsymbol{LFI:white:noise:sensitivity:LFI26:Rad:M}{478.6}
\setsymbol{LFI:white:noise:sensitivity:LFI27:Rad:M}{277.7}
\setsymbol{LFI:white:noise:sensitivity:LFI28:Rad:M}{312.3}
\setsymbol{LFI:white:noise:sensitivity:LFI18:Rad:S}{465.7}
\setsymbol{LFI:white:noise:sensitivity:LFI19:Rad:S}{555.6}
\setsymbol{LFI:white:noise:sensitivity:LFI20:Rad:S}{623.2}
\setsymbol{LFI:white:noise:sensitivity:LFI21:Rad:S}{564.1}
\setsymbol{LFI:white:noise:sensitivity:LFI22:Rad:S}{534.4}
\setsymbol{LFI:white:noise:sensitivity:LFI23:Rad:S}{542.4}
\setsymbol{LFI:white:noise:sensitivity:LFI24:Rad:S}{399.2}
\setsymbol{LFI:white:noise:sensitivity:LFI25:Rad:S}{392.6}
\setsymbol{LFI:white:noise:sensitivity:LFI26:Rad:S}{418.6}
\setsymbol{LFI:white:noise:sensitivity:LFI27:Rad:S}{302.9}
\setsymbol{LFI:white:noise:sensitivity:LFI28:Rad:S}{285.3}


\setsymbol{LFI:knee:frequency:70GHz:units}{29.5\mHz}
\setsymbol{LFI:knee:frequency:44GHz:units}{56.2\mHz}
\setsymbol{LFI:knee:frequency:30GHz:units}{113.7\mHz}

\setsymbol{LFI:knee:frequency:70GHz}{29.5}
\setsymbol{LFI:knee:frequency:44GHz}{56.2}
\setsymbol{LFI:knee:frequency:30GHz}{113.7}

\setsymbol{LFI:knee:frequency:LFI18:Rad:M:units}{16.3\mHz}
\setsymbol{LFI:knee:frequency:LFI19:Rad:M:units}{15.1\mHz}
\setsymbol{LFI:knee:frequency:LFI20:Rad:M:units}{18.7\mHz}
\setsymbol{LFI:knee:frequency:LFI21:Rad:M:units}{37.2\mHz}
\setsymbol{LFI:knee:frequency:LFI22:Rad:M:units}{12.7\mHz}
\setsymbol{LFI:knee:frequency:LFI23:Rad:M:units}{34.6\mHz}
\setsymbol{LFI:knee:frequency:LFI24:Rad:M:units}{46.2\mHz}
\setsymbol{LFI:knee:frequency:LFI25:Rad:M:units}{24.9\mHz}
\setsymbol{LFI:knee:frequency:LFI26:Rad:M:units}{67.6\mHz}
\setsymbol{LFI:knee:frequency:LFI27:Rad:M:units}{187.4\mHz}
\setsymbol{LFI:knee:frequency:LFI28:Rad:M:units}{122.2\mHz}
\setsymbol{LFI:knee:frequency:LFI18:Rad:S:units}{17.7\mHz}
\setsymbol{LFI:knee:frequency:LFI19:Rad:S:units}{22.0\mHz}
\setsymbol{LFI:knee:frequency:LFI20:Rad:S:units}{8.7\mHz}
\setsymbol{LFI:knee:frequency:LFI21:Rad:S:units}{25.9\mHz}
\setsymbol{LFI:knee:frequency:LFI22:Rad:S:units}{15.8\mHz}
\setsymbol{LFI:knee:frequency:LFI23:Rad:S:units}{129.8\mHz}
\setsymbol{LFI:knee:frequency:LFI24:Rad:S:units}{100.9\mHz}
\setsymbol{LFI:knee:frequency:LFI25:Rad:S:units}{38.9\mHz}
\setsymbol{LFI:knee:frequency:LFI26:Rad:S:units}{58.9\mHz}
\setsymbol{LFI:knee:frequency:LFI27:Rad:S:units}{104.4\mHz}
\setsymbol{LFI:knee:frequency:LFI28:Rad:S:units}{40.7\mHz}

\setsymbol{LFI:knee:frequency:LFI18:Rad:M}{16.3}
\setsymbol{LFI:knee:frequency:LFI19:Rad:M}{15.1}
\setsymbol{LFI:knee:frequency:LFI20:Rad:M}{18.7}
\setsymbol{LFI:knee:frequency:LFI21:Rad:M}{37.2}
\setsymbol{LFI:knee:frequency:LFI22:Rad:M}{12.7}
\setsymbol{LFI:knee:frequency:LFI23:Rad:M}{34.6}
\setsymbol{LFI:knee:frequency:LFI24:Rad:M}{46.2}
\setsymbol{LFI:knee:frequency:LFI25:Rad:M}{24.9}
\setsymbol{LFI:knee:frequency:LFI26:Rad:M}{67.6}
\setsymbol{LFI:knee:frequency:LFI27:Rad:M}{187.4}
\setsymbol{LFI:knee:frequency:LFI28:Rad:M}{122.2}
\setsymbol{LFI:knee:frequency:LFI18:Rad:S}{17.7}
\setsymbol{LFI:knee:frequency:LFI19:Rad:S}{22.0}
\setsymbol{LFI:knee:frequency:LFI20:Rad:S}{8.7}
\setsymbol{LFI:knee:frequency:LFI21:Rad:S}{25.9}
\setsymbol{LFI:knee:frequency:LFI22:Rad:S}{15.8}
\setsymbol{LFI:knee:frequency:LFI23:Rad:S}{129.8}
\setsymbol{LFI:knee:frequency:LFI24:Rad:S}{100.9}
\setsymbol{LFI:knee:frequency:LFI25:Rad:S}{38.9}
\setsymbol{LFI:knee:frequency:LFI26:Rad:S}{58.9}
\setsymbol{LFI:knee:frequency:LFI27:Rad:S}{104.4}
\setsymbol{LFI:knee:frequency:LFI28:Rad:S}{40.7}


\setsymbol{LFI:slope:70GHz:units}{$-1.03$\mHz}
\setsymbol{LFI:slope:44GHz:units}{$-0.89$\mHz}
\setsymbol{LFI:slope:30GHz:units}{$-0.87$\mHz}

\setsymbol{LFI:slope:70GHz}{$-1.03$}
\setsymbol{LFI:slope:44GHz}{$-0.89$}
\setsymbol{LFI:slope:30GHz}{$-0.87$}

\setsymbol{LFI:slope:LFI18:Rad:M:units}{$-1.04$\mHz}
\setsymbol{LFI:slope:LFI19:Rad:M:units}{$-1.09$\mHz}
\setsymbol{LFI:slope:LFI20:Rad:M:units}{$-0.69$\mHz}
\setsymbol{LFI:slope:LFI21:Rad:M:units}{$-1.56$\mHz}
\setsymbol{LFI:slope:LFI22:Rad:M:units}{$-1.01$\mHz}
\setsymbol{LFI:slope:LFI23:Rad:M:units}{$-0.96$\mHz}
\setsymbol{LFI:slope:LFI24:Rad:M:units}{$-0.83$\mHz}
\setsymbol{LFI:slope:LFI25:Rad:M:units}{$-0.91$\mHz}
\setsymbol{LFI:slope:LFI26:Rad:M:units}{$-0.95$\mHz}
\setsymbol{LFI:slope:LFI27:Rad:M:units}{$-0.87$\mHz}
\setsymbol{LFI:slope:LFI28:Rad:M:units}{$-0.88$\mHz}
\setsymbol{LFI:slope:LFI18:Rad:S:units}{$-1.15$\mHz}
\setsymbol{LFI:slope:LFI19:Rad:S:units}{$-1.00$\mHz}
\setsymbol{LFI:slope:LFI20:Rad:S:units}{$-0.95$\mHz}
\setsymbol{LFI:slope:LFI21:Rad:S:units}{$-0.92$\mHz}
\setsymbol{LFI:slope:LFI22:Rad:S:units}{$-1.01$\mHz}
\setsymbol{LFI:slope:LFI23:Rad:S:units}{$-0.95$\mHz}
\setsymbol{LFI:slope:LFI24:Rad:S:units}{$-0.73$\mHz}
\setsymbol{LFI:slope:LFI25:Rad:S:units}{$-1.16$\mHz}
\setsymbol{LFI:slope:LFI26:Rad:S:units}{$-0.79$\mHz}
\setsymbol{LFI:slope:LFI27:Rad:S:units}{$-0.82$\mHz}
\setsymbol{LFI:slope:LFI28:Rad:S:units}{$-0.91$\mHz}

\setsymbol{LFI:slope:LFI18:Rad:M}{$-1.04$}
\setsymbol{LFI:slope:LFI19:Rad:M}{$-1.09$}
\setsymbol{LFI:slope:LFI20:Rad:M}{$-0.69$}
\setsymbol{LFI:slope:LFI21:Rad:M}{$-1.56$}
\setsymbol{LFI:slope:LFI22:Rad:M}{$-1.01$}
\setsymbol{LFI:slope:LFI23:Rad:M}{$-0.96$}
\setsymbol{LFI:slope:LFI24:Rad:M}{$-0.83$}
\setsymbol{LFI:slope:LFI25:Rad:M}{$-0.91$}
\setsymbol{LFI:slope:LFI26:Rad:M}{$-0.95$}
\setsymbol{LFI:slope:LFI27:Rad:M}{$-0.87$}
\setsymbol{LFI:slope:LFI28:Rad:M}{$-0.88$}
\setsymbol{LFI:slope:LFI18:Rad:S}{$-1.15$}
\setsymbol{LFI:slope:LFI19:Rad:S}{$-1.00$}
\setsymbol{LFI:slope:LFI20:Rad:S}{$-0.95$}
\setsymbol{LFI:slope:LFI21:Rad:S}{$-0.92$}
\setsymbol{LFI:slope:LFI22:Rad:S}{$-1.01$}
\setsymbol{LFI:slope:LFI23:Rad:S}{$-0.95$}
\setsymbol{LFI:slope:LFI24:Rad:S}{$-0.73$}
\setsymbol{LFI:slope:LFI25:Rad:S}{$-1.16$}
\setsymbol{LFI:slope:LFI26:Rad:S}{$-0.79$}
\setsymbol{LFI:slope:LFI27:Rad:S}{$-0.82$}
\setsymbol{LFI:slope:LFI28:Rad:S}{$-0.91$}


\setsymbol{LFI:FWHM:70GHz:units}{13\parcm01}
\setsymbol{LFI:FWHM:44GHz:units}{27\parcm92}
\setsymbol{LFI:FWHM:30GHz:units}{32\parcm65}

\setsymbol{LFI:FWHM:70GHz}{13.01}
\setsymbol{LFI:FWHM:44GHz}{27.92}
\setsymbol{LFI:FWHM:30GHz}{32.65}

\setsymbol{LFI:FWHM:LFI18:units}{13\parcm39}
\setsymbol{LFI:FWHM:LFI19:units}{13\parcm01}
\setsymbol{LFI:FWHM:LFI20:units}{12\parcm75}
\setsymbol{LFI:FWHM:LFI21:units}{12\parcm74}
\setsymbol{LFI:FWHM:LFI22:units}{12\parcm87}
\setsymbol{LFI:FWHM:LFI23:units}{13\parcm27}
\setsymbol{LFI:FWHM:LFI24:units}{22\parcm98}
\setsymbol{LFI:FWHM:LFI25:units}{30\parcm46}
\setsymbol{LFI:FWHM:LFI26:units}{30\parcm31}
\setsymbol{LFI:FWHM:LFI27:units}{32\parcm65}
\setsymbol{LFI:FWHM:LFI28:units}{32\parcm66}

\setsymbol{LFI:FWHM:LFI18}{13.39}
\setsymbol{LFI:FWHM:LFI19}{13.01}
\setsymbol{LFI:FWHM:LFI20}{12.75}
\setsymbol{LFI:FWHM:LFI21}{12.74}
\setsymbol{LFI:FWHM:LFI22}{12.87}
\setsymbol{LFI:FWHM:LFI23}{13.27}
\setsymbol{LFI:FWHM:LFI24}{22.98}
\setsymbol{LFI:FWHM:LFI25}{30.46}
\setsymbol{LFI:FWHM:LFI26}{30.31}
\setsymbol{LFI:FWHM:LFI27}{32.65}
\setsymbol{LFI:FWHM:LFI28}{32.66}



\setsymbol{LFI:FWHM:uncertainty:LFI18:units}{0.170\arcm}
\setsymbol{LFI:FWHM:uncertainty:LFI19:units}{0.174\arcm}
\setsymbol{LFI:FWHM:uncertainty:LFI20:units}{0.170\arcm}
\setsymbol{LFI:FWHM:uncertainty:LFI21:units}{0.156\arcm}
\setsymbol{LFI:FWHM:uncertainty:LFI22:units}{0.164\arcm}
\setsymbol{LFI:FWHM:uncertainty:LFI23:units}{0.171\arcm}
\setsymbol{LFI:FWHM:uncertainty:LFI24:units}{0.652\arcm}
\setsymbol{LFI:FWHM:uncertainty:LFI25:units}{1.075\arcm}
\setsymbol{LFI:FWHM:uncertainty:LFI26:units}{1.131\arcm}
\setsymbol{LFI:FWHM:uncertainty:LFI27:units}{1.266\arcm}
\setsymbol{LFI:FWHM:uncertainty:LFI28:units}{1.287\arcm}

\setsymbol{LFI:FWHM:uncertainty:LFI18}{0.170}
\setsymbol{LFI:FWHM:uncertainty:LFI19}{0.174}
\setsymbol{LFI:FWHM:uncertainty:LFI20}{0.170}
\setsymbol{LFI:FWHM:uncertainty:LFI21}{0.156}
\setsymbol{LFI:FWHM:uncertainty:LFI22}{0.164}
\setsymbol{LFI:FWHM:uncertainty:LFI23}{0.171}
\setsymbol{LFI:FWHM:uncertainty:LFI24}{0.652}
\setsymbol{LFI:FWHM:uncertainty:LFI25}{1.075}
\setsymbol{LFI:FWHM:uncertainty:LFI26}{1.131}
\setsymbol{LFI:FWHM:uncertainty:LFI27}{1.266}
\setsymbol{LFI:FWHM:uncertainty:LFI28}{1.287}


\setsymbol{HFI:center:frequency:100GHz:units}{100\,GHz}
\setsymbol{HFI:center:frequency:143GHz:units}{143\,GHz}
\setsymbol{HFI:center:frequency:217GHz:units}{217\,GHz}
\setsymbol{HFI:center:frequency:353GHz:units}{353\,GHz}
\setsymbol{HFI:center:frequency:545GHz:units}{545\,GHz}
\setsymbol{HFI:center:frequency:857GHz:units}{857\,GHz}

\setsymbol{HFI:center:frequency:100GHz}{100}
\setsymbol{HFI:center:frequency:143GHz}{143}
\setsymbol{HFI:center:frequency:217GHz}{217}
\setsymbol{HFI:center:frequency:353GHz}{353}
\setsymbol{HFI:center:frequency:545GHz}{545}
\setsymbol{HFI:center:frequency:857GHz}{857}


\setsymbol{HFI:Ndetectors:100GHz}{8}
\setsymbol{HFI:Ndetectors:143GHz}{11}
\setsymbol{HFI:Ndetectors:217GHz}{12}
\setsymbol{HFI:Ndetectors:353GHz}{12}
\setsymbol{HFI:Ndetectors:545GHz}{3}
\setsymbol{HFI:Ndetectors:857GHz}{4}


\setsymbol{HFI:FWHM:Maps:100GHz:units}{9\parcm88}
\setsymbol{HFI:FWHM:Maps:143GHz:units}{7\parcm18}
\setsymbol{HFI:FWHM:Maps:217GHz:units}{4\parcm87}
\setsymbol{HFI:FWHM:Maps:353GHz:units}{4\parcm65}
\setsymbol{HFI:FWHM:Maps:545GHz:units}{4\parcm72}
\setsymbol{HFI:FWHM:Maps:857GHz:units}{4\parcm39}
\setsymbol{HFI:FWHM:Maps:100GHz}{9.88}
\setsymbol{HFI:FWHM:Maps:143GHz}{7.18}
\setsymbol{HFI:FWHM:Maps:217GHz}{4.87}
\setsymbol{HFI:FWHM:Maps:353GHz}{4.65}
\setsymbol{HFI:FWHM:Maps:545GHz}{4.72}
\setsymbol{HFI:FWHM:Maps:857GHz}{4.39}


\setsymbol{HFI:beam:ellipticity:Maps:100GHz}{1.15}
\setsymbol{HFI:beam:ellipticity:Maps:143GHz}{1.01}
\setsymbol{HFI:beam:ellipticity:Maps:217GHz}{1.06}
\setsymbol{HFI:beam:ellipticity:Maps:353GHz}{1.05}
\setsymbol{HFI:beam:ellipticity:Maps:545GHz}{1.14}
\setsymbol{HFI:beam:ellipticity:Maps:857GHz}{1.19}


\setsymbol{HFI:FWHM:Mars:100GHz:units}{9\parcm37}
\setsymbol{HFI:FWHM:Mars:143GHz:units}{7\parcm04}
\setsymbol{HFI:FWHM:Mars:217GHz:units}{4\parcm68}
\setsymbol{HFI:FWHM:Mars:353GHz:units}{4\parcm43}
\setsymbol{HFI:FWHM:Mars:545GHz:units}{3\parcm80}
\setsymbol{HFI:FWHM:Mars:857GHz:units}{3\parcm67}

\setsymbol{HFI:FWHM:Mars:100GHz}{9.37}
\setsymbol{HFI:FWHM:Mars:143GHz}{7.04}
\setsymbol{HFI:FWHM:Mars:217GHz}{4.68}
\setsymbol{HFI:FWHM:Mars:353GHz}{4.43}
\setsymbol{HFI:FWHM:Mars:545GHz}{3.80}
\setsymbol{HFI:FWHM:Mars:857GHz}{3.67}


\setsymbol{HFI:beam:ellipticity:Mars:100GHz}{1.18}
\setsymbol{HFI:beam:ellipticity:Mars:143GHz}{1.03}
\setsymbol{HFI:beam:ellipticity:Mars:217GHz}{1.14}
\setsymbol{HFI:beam:ellipticity:Mars:353GHz}{1.09}
\setsymbol{HFI:beam:ellipticity:Mars:545GHz}{1.25}
\setsymbol{HFI:beam:ellipticity:Mars:857GHz}{1.03}


\setsymbol{HFI:CMB:relative:calibration:100GHz}{$\lsim 1\%$}
\setsymbol{HFI:CMB:relative:calibration:143GHz}{$\lsim 1\%$}
\setsymbol{HFI:CMB:relative:calibration:217GHz}{$\lsim 1\%$}
\setsymbol{HFI:CMB:relative:calibration:353GHz}{$\lsim 1\%$}
\setsymbol{HFI:CMB:relative:calibration:545GHz}{}
\setsymbol{HFI:CMB:relative:calibration:857GHz}{}


\setsymbol{HFI:CMB:absolute:calibration:100GHz}{$\lsim 2\%$}
\setsymbol{HFI:CMB:absolute:calibration:143GHz}{$\lsim 2\%$}
\setsymbol{HFI:CMB:absolute:calibration:217GHz}{$\lsim 2\%$}
\setsymbol{HFI:CMB:absolute:calibration:353GHz}{$\lsim 2\%$}
\setsymbol{HFI:CMB:absolute:calibration:545GHz}{}
\setsymbol{HFI:CMB:absolute:calibration:857GHz}{}


\setsymbol{HFI:FIRAS:gain:calibration:accuracy:statistical:100GHz}{}
\setsymbol{HFI:FIRAS:gain:calibration:accuracy:statistical:143GHz}{}
\setsymbol{HFI:FIRAS:gain:calibration:accuracy:statistical:217GHz}{}
\setsymbol{HFI:FIRAS:gain:calibration:accuracy:statistical:353GHz}{2.5\%}
\setsymbol{HFI:FIRAS:gain:calibration:accuracy:statistical:545GHz}{1\%}
\setsymbol{HFI:FIRAS:gain:calibration:accuracy:statistical:857GHz}{0.5\%}


\setsymbol{HFI:FIRAS:gain:calibration:accuracy:systematic:100GHz}{}
\setsymbol{HFI:FIRAS:gain:calibration:accuracy:systematic:143GHz}{}
\setsymbol{HFI:FIRAS:gain:calibration:accuracy:systematic:217GHz}{}
\setsymbol{HFI:FIRAS:gain:calibration:accuracy:systematic:353GHz}{}
\setsymbol{HFI:FIRAS:gain:calibration:accuracy:systematic:545GHz}{7\%}
\setsymbol{HFI:FIRAS:gain:calibration:accuracy:systematic:857GHz}{7\%}


\setsymbol{HFI:FIRAS:zero:point:accuracy:100GHz:units}{0.8\MJysr}
\setsymbol{HFI:FIRAS:zero:point:accuracy:143GHz:units}{}
\setsymbol{HFI:FIRAS:zero:point:accuracy:217GHz:units}{}
\setsymbol{HFI:FIRAS:zero:point:accuracy:353GHz:units}{1.4\MJysr}
\setsymbol{HFI:FIRAS:zero:point:accuracy:545GHz:units}{2.2\MJysr}
\setsymbol{HFI:FIRAS:zero:point:accuracy:857GHz:units}{1.7\MJysr}

\setsymbol{HFI:FIRAS:zero:point:accuracy:100GHz}{0.8}
\setsymbol{HFI:FIRAS:zero:point:accuracy:143GHz}{}
\setsymbol{HFI:FIRAS:zero:point:accuracy:217GHz}{}
\setsymbol{HFI:FIRAS:zero:point:accuracy:353GHz}{1.4}
\setsymbol{HFI:FIRAS:zero:point:accuracy:545GHz}{2.2}
\setsymbol{HFI:FIRAS:zero:point:accuracy:857GHz}{1.7}


\setsymbol{HFI:unit:conversion:100GHz:units}{0.2415\MJysrmK}
\setsymbol{HFI:unit:conversion:143GHz:units}{0.3694\MJysrmK}
\setsymbol{HFI:unit:conversion:217GHz:units}{0.4811\MJysrmK}
\setsymbol{HFI:unit:conversion:353GHz:units}{0.2883\MJysrmK}
\setsymbol{HFI:unit:conversion:545GHz:units}{0.05826\MJysrmK}
\setsymbol{HFI:unit:conversion:857GHz:units}{0.002238\MJysrmK}

\setsymbol{HFI:unit:conversion:100GHz}{0.2415}
\setsymbol{HFI:unit:conversion:143GHz}{0.3694}
\setsymbol{HFI:unit:conversion:217GHz}{0.4811}
\setsymbol{HFI:unit:conversion:353GHz}{0.2883}
\setsymbol{HFI:unit:conversion:545GHz}{0.05826}
\setsymbol{HFI:unit:conversion:857GHz}{0.002238}


\setsymbol{HFI:colour:correction:alpha=-2:V1.01:100GHz}{0.9893}
\setsymbol{HFI:colour:correction:alpha=-2:V1.01:143GHz}{0.9759}
\setsymbol{HFI:colour:correction:alpha=-2:V1.01:217GHz}{1.0007}
\setsymbol{HFI:colour:correction:alpha=-2:V1.01:353GHz}{1.0028}
\setsymbol{HFI:colour:correction:alpha=-2:V1.01:545GHz}{1.0019}
\setsymbol{HFI:colour:correction:alpha=-2:V1.01:857GHz}{0.9889}


\setsymbol{HFI:colour:correction:alpha=0:V1.01:100GHz}{1.0008}
\setsymbol{HFI:colour:correction:alpha=0:V1.01:143GHz}{1.0148}
\setsymbol{HFI:colour:correction:alpha=0:V1.01:217GHz}{0.9909}
\setsymbol{HFI:colour:correction:alpha=0:V1.01:353GHz}{0.9888}
\setsymbol{HFI:colour:correction:alpha=0:V1.01:545GHz}{0.9878}
\setsymbol{HFI:colour:correction:alpha=0:V1.01:857GHz}{1.0014}

\usepackage{graphicx}
\usepackage{color}
\usepackage{natbib}
\usepackage{longtable,lscape}
\usepackage{txfonts}
\usepackage{amssymb}
\usepackage{amsfonts}
\usepackage{url}
\usepackage[breaklinks, colorlinks, citecolor=blue]{hyperref}
\usepackage{booktabs}
\usepackage{subfig}
\usepackage{sidecap}

\def\reff@jnl#1{{#1\/}}
\def\apj{\reff@jnl{ApJ}}       
\def\apjs{\reff@jnl{ApJS}}     
\def\aaps{\reff@jnl{A\&AS}}    
\def\mnras{\reff@jnl{MNRAS}}   
\def\prd{\reff@jnl{Phys.\ Rev.\ D}}    



\newcommand{\beq}{\begin{equation}}
\newcommand{\eeq}{\end{equation}}

\newcommand{\be}{\begin{equation}}
\newcommand{\ee}{\end{equation}}
\newcommand{\bea}{\begin{eq}}
\newcommand{\eea}{\end{equation}}

{\begin{enumerate}\setlength{\itemsep}{0mm}}%
{\end{enumerate}}
{\begin{enumerate}\setlength{\itemsep}{0mm}}%
{\end{enumerate}}

\newcommand{\bc}{\begin{center}}
\newcommand{\ec}{\end{center}}
\newcommand{\bi}{\begin{itemize}}
\newcommand{\ei}{\end{itemize}}
\newcommand{\ben}{\begin{enumerate}}
\newcommand{\een}{\end{enumerate}}

\newfont{\gwpfont}{cmssq8 scaled 1000}

\def\msol {\mathrm{M}_{\odot}}
\def\YX {Y_\mathrm{X}}

\def\YSZ{Y_\mathrm{500}}

\def\YSZ {Y_{500}}
\def\Mv {M_{500}}

\def\tv {\theta_{500}}

\def\da {D_{\mathrm A}^{2}}
\def\MYSZ {$\Mv$--$\da\YSZ$}

\def\lesssim{\mathrel{\hbox{\rlap{\hbox{\lower4pt\hbox{$\sim$}}}\hbox{$<$}}}}
\def\gtrsim{\mathrel{\hbox{\rlap{\hbox{\lower4pt\hbox{$\sim$}}}\hbox{$>$}}}}

\newcommand{\propsim}{\lower 3pt \hbox{$\, \buildrel {\textstyle
     \propto}\over {\textstyle \sim}\,$}}

\citestyle{aa}



\def\planck{\Planck}

\def\lesssim{\mathrel{\hbox{\rlap{\hbox{\lower4pt\hbox{$\sim$}}}\hbox{$<$}}}}
\def\gtrsim{\mathrel{\hbox{\rlap{\hbox{\lower4pt\hbox{$\sim$}}}\hbox{$>$}}}}


\begin{document}
\author{\small
Planck Collaboration:
P.~A.~R.~Ade\inst{99}
\and
N.~Aghanim\inst{68}\thanks{Corresponding author: N.~Aghanim $\;\;\;\;\;\;\;\; $\url{nabila.aghanim@ias.u-psud.fr}}
\and
C.~Armitage-Caplan\inst{104}
\and
M.~Arnaud\inst{81}
\and
M.~Ashdown\inst{78, 7}
\and
F.~Atrio-Barandela\inst{21}
\and
J.~Aumont\inst{68}
\and
H.~Aussel\inst{81}
\and
C.~Baccigalupi\inst{97}
% ...
}
\institute{\small
APC, AstroParticule et Cosmologie, Universit\'{e} Paris Diderot, CNRS/IN2P3, CEA/lrfu, Observatoire de Paris, Sorbonne Paris Cit\'{e}, 10, rue Alice Domon et L\'{e}onie Duquet, 75205 Paris Cedex 13, France\\
\and
Aalto University Mets\"{a}hovi Radio Observatory, Mets\"{a}hovintie 114, FIN-02540 Kylm\"{a}l\"{a}, Finland\\
\and
Academy of Sciences of Tatarstan, Bauman Str., 20, Kazan, 420111, Republic of Tatarstan, Russia\\
\and
African Institute for Mathematical Sciences, 6-8 Melrose Road, Muizenberg, Cape Town, South Africa\\
\and
Agenzia Spaziale Italiana Science Data Center, Via del Politecnico snc, 00133, Roma, Italy\\
\and
Agenzia Spaziale Italiana, Viale Liegi 26, Roma, Italy\\
\and
Astrophysics Group, Cavendish Laboratory, University of Cambridge, J J Thomson Avenue, Cambridge CB3 0HE, U.K.\\
\and
Astrophysics \& Cosmology Research Unit, School of Mathematics, Statistics \& Computer Science, University of KwaZulu-Natal, Westville Campus, Private Bag X54001, Durban 4000, South Africa\\
\and
Atacama Large Millimeter/submillimeter Array, ALMA Santiago Central Offices, Alonso de Cordova 3107, Vitacura, Casilla 763 0355, Santiago, Chile\\
\and
CITA, University of Toronto, 60 St. George St., Toronto, ON M5S 3H8, Canada\\
\and
CNRS, IRAP, 9 Av. colonel Roche, BP 44346, F-31028 Toulouse cedex 4, France\\
\and
California Institute of Technology, Pasadena, California, U.S.A.\\
\and
Centre for Theoretical Cosmology, DAMTP, University of Cambridge, Wilberforce Road, Cambridge CB3 0WA, U.K.\\
\and
Centro de Astrof\'{\i}sica, Universidade do Porto, Rua das Estrelas, 4150-762 Porto, Portugal\\
\and
Centro de Estudios de F\'{i}sica del Cosmos de Arag\'{o}n (CEFCA), Plaza San Juan, 1, planta 2, E-44001, Teruel, Spain\\
\and
Computational Cosmology Center, Lawrence Berkeley National Laboratory, Berkeley, California, U.S.A.\\
\and
Consejo Superior de Investigaciones Cient\'{\i}ficas (CSIC), Madrid, Spain\\
\and
DSM/Irfu/SPP, CEA-Saclay, F-91191 Gif-sur-Yvette Cedex, France\\
\and
DTU Space, National Space Institute, Technical University of Denmark, Elektrovej 327, DK-2800 Kgs. Lyngby, Denmark\\
\and
D\'{e}partement de Physique Th\'{e}orique, Universit\'{e} de Gen\`{e}ve, 24, Quai E. Ansermet,1211 Gen\`{e}ve 4, Switzerland\\
\and
Departamento de F\'{\i}sica Fundamental, Facultad de Ciencias, Universidad de Salamanca, 37008 Salamanca, Spain\\
\and
Departamento de F\'{\i}sica, Universidad de Oviedo, Avda. Calvo Sotelo s/n, Oviedo, Spain\\
\and
Department of Astronomy and Astrophysics, University of Toronto, 50 Saint George Street, Toronto, Ontario, Canada\\
\and
Department of Astronomy and Geodesy, Kazan Federal University,  Kremlevskaya Str., 18, Kazan, 420008, Russia\\
\and
Department of Astrophysics/IMAPP, Radboud University Nijmegen, P.O. Box 9010, 6500 GL Nijmegen, The Netherlands\\
\and
Department of Electrical Engineering and Computer Sciences, University of California, Berkeley, California, U.S.A.\\
\and
Department of Physics \& Astronomy, University of British Columbia, 6224 Agricultural Road, Vancouver, British Columbia, Canada\\
\and
Department of Physics and Astronomy, Dana and David Dornsife College of Letter, Arts and Sciences, University of Southern California, Los Angeles, CA 90089, U.S.A.\\
\and
Department of Physics and Astronomy, University College London, London WC1E 6BT, U.K.\\
\and
Department of Physics and Astronomy, University of Sussex, Brighton BN1 9QH, U.K.\\
\and
Department of Physics, Florida State University, Keen Physics Building, 77 Chieftan Way, Tallahassee, Florida, U.S.A.\\
\and
Department of Physics, Gustaf H\"{a}llstr\"{o}min katu 2a, University of Helsinki, Helsinki, Finland\\
\and
Department of Physics, Princeton University, Princeton, New Jersey, U.S.A.\\
\and
Department of Physics, University of California, Berkeley, California, U.S.A.\\
\and
Department of Physics, University of California, One Shields Avenue, Davis, California, U.S.A.\\
\and
Department of Physics, University of California, Santa Barbara, California, U.S.A.\\
\and
Department of Physics, University of Illinois at Urbana-Champaign, 1110 West Green Street, Urbana, Illinois, U.S.A.\\
\and
Dipartimento di Fisica e Astronomia G. Galilei, Universit\`{a} degli Studi di Padova, via Marzolo 8, 35131 Padova, Italy\\
\and
Dipartimento di Fisica e Scienze della Terra, Universit\`{a} di Ferrara, Via Saragat 1, 44122 Ferrara, Italy\\
\and
Dipartimento di Fisica, Universit\`{a} La Sapienza, P. le A. Moro 2, Roma, Italy\\
\and
Dipartimento di Fisica, Universit\`{a} degli Studi di Milano, Via Celoria, 16, Milano, Italy\\
\and
Dipartimento di Fisica, Universit\`{a} degli Studi di Trieste, via A. Valerio 2, Trieste, Italy\\
\and
Dipartimento di Fisica, Universit\`{a} di Roma Tor Vergata, Via della Ricerca Scientifica, 1, Roma, Italy\\
\and
Discovery Center, Niels Bohr Institute, Blegdamsvej 17, Copenhagen, Denmark\\
\and
Dpto. Astrof\'{i}sica, Universidad de La Laguna (ULL), E-38206 La Laguna, Tenerife, Spain\\
\and
European Southern Observatory, ESO Vitacura, Alonso de Cordova 3107, Vitacura, Casilla 19001, Santiago, Chile\\
\and
European Space Agency, ESAC, Planck Science Office, Camino bajo del Castillo, s/n, Urbanizaci\'{o}n Villafranca del Castillo, Villanueva de la Ca\~{n}ada, Madrid, Spain\\
\and
European Space Agency, ESTEC, Keplerlaan 1, 2201 AZ Noordwijk, The Netherlands\\
\and
Finnish Centre for Astronomy with ESO (FINCA), University of Turku, V\"{a}is\"{a}l\"{a}ntie 20, FIN-21500, Piikki\"{o}, Finland\\
\and
GEPI, Observatoire de Paris, Section de Meudon, 5 Place J. Janssen, 92195 Meudon Cedex, France\\
\and
Helsinki Institute of Physics, Gustaf H\"{a}llstr\"{o}min katu 2, University of Helsinki, Helsinki, Finland\\
\and
INAF - Osservatorio Astrofisico di Catania, Via S. Sofia 78, Catania, Italy\\
\and
INAF - Osservatorio Astronomico di Padova, Vicolo dell'Osservatorio 5, Padova, Italy\\
\and
INAF - Osservatorio Astronomico di Roma, via di Frascati 33, Monte Porzio Catone, Italy\\
\and
INAF - Osservatorio Astronomico di Trieste, Via G.B. Tiepolo 11, Trieste, Italy\\
\and
INAF Istituto di Radioastronomia, Via P. Gobetti 101, 40129 Bologna, Italy\\
\and
INAF/IASF Bologna, Via Gobetti 101, Bologna, Italy\\
\and
INAF/IASF Milano, Via E. Bassini 15, Milano, Italy\\
\and
INFN, Sezione di Bologna, Via Irnerio 46, I-40126, Bologna, Italy\\
\and
INFN, Sezione di Roma 1, Universit\`{a} di Roma Sapienza, Piazzale Aldo Moro 2, 00185, Roma, Italy\\
\and
IPAG: Institut de Plan\'{e}tologie et d'Astrophysique de Grenoble, Universit\'{e} Joseph Fourier, Grenoble 1 / CNRS-INSU, UMR 5274, Grenoble, F-38041, France\\
\and
ISDC Data Centre for Astrophysics, University of Geneva, ch. d'Ecogia 16, Versoix, Switzerland\\
\and
IUCAA, Post Bag 4, Ganeshkhind, Pune University Campus, Pune 411 007, India\\
\and
Imperial College London, Astrophysics group, Blackett Laboratory, Prince Consort Road, London, SW7 2AZ, U.K.\\
\and
Infrared Processing and Analysis Center, California Institute of Technology, Pasadena, CA 91125, U.S.A.\\
\and
Institut N\'{e}el, CNRS, Universit\'{e} Joseph Fourier Grenoble I, 25 rue des Martyrs, Grenoble, France\\
\and
Institut Universitaire de France, 103, bd Saint-Michel, 75005, Paris, France\\
\and
Institut d'Astrophysique Spatiale, CNRS (UMR8617) Universit\'{e} Paris-Sud 11, B\^{a}timent 121, Orsay, France\\
\and
Institut d'Astrophysique de Paris, CNRS (UMR7095), 98 bis Boulevard Arago, F-75014, Paris, France\\
\and
Institute for Space Sciences, Bucharest-Magurale, Romania\\
\and
Institute of Astronomy and Astrophysics, Academia Sinica, Taipei, Taiwan\\
\and
Institute of Astronomy, University of Cambridge, Madingley Road, Cambridge CB3 0HA, U.K.\\
\and
Institute of Theoretical Astrophysics, University of Oslo, Blindern, Oslo, Norway\\
\and
Instituto de Astrof\'{\i}sica de Canarias, C/V\'{\i}a L\'{a}ctea s/n, La Laguna, Tenerife, Spain\\
\and
Instituto de F\'{\i}sica de Cantabria (CSIC-Universidad de Cantabria), Avda. de los Castros s/n, Santander, Spain\\
\and
Jet Propulsion Laboratory, California Institute of Technology, 4800 Oak Grove Drive, Pasadena, California, U.S.A.\\
\and
Jodrell Bank Centre for Astrophysics, Alan Turing Building, School of Physics and Astronomy, The University of Manchester, Oxford Road, Manchester, M13 9PL, U.K.\\
\and
Kavli Institute for Cosmology Cambridge, Madingley Road, Cambridge, CB3 0HA, U.K.\\
\and
LAL, Universit\'{e} Paris-Sud, CNRS/IN2P3, Orsay, France\\
\and
LERMA, CNRS, Observatoire de Paris, 61 Avenue de l'Observatoire, Paris, France\\
\and
Laboratoire AIM, IRFU/Service d'Astrophysique - CEA/DSM - CNRS - Universit\'{e} Paris Diderot, B\^{a}t. 709, CEA-Saclay, F-91191 Gif-sur-Yvette Cedex, France\\
\and
Laboratoire Traitement et Communication de l'Information, CNRS (UMR 5141) and T\'{e}l\'{e}com ParisTech, 46 rue Barrault F-75634 Paris Cedex 13, France\\
\and
Laboratoire de Physique Subatomique et de Cosmologie, Universit\'{e} Joseph Fourier Grenoble I, CNRS/IN2P3, Institut National Polytechnique de Grenoble, 53 rue des Martyrs, 38026 Grenoble cedex, France\\
\and
Laboratoire de Physique Th\'{e}orique, Universit\'{e} Paris-Sud 11 \& CNRS, B\^{a}timent 210, 91405 Orsay, France\\
\and
Lawrence Berkeley National Laboratory, Berkeley, California, U.S.A.\\
\and
MPA Partner Group, Key Laboratory for Research in Galaxies and Cosmology, Shanghai Astronomical Observatory, Chinese Academy of Sciences, Nandan Road 80, Shanghai 200030, China\\
\and
Max-Planck-Institut f\"{u}r Astrophysik, Karl-Schwarzschild-Str. 1, 85741 Garching, Germany\\
\and
Max-Planck-Institut f\"{u}r Extraterrestrische Physik, Giessenbachstra{\ss}e, 85748 Garching, Germany\\
\and
McGill Physics, Ernest Rutherford Physics Building, McGill University, 3600 rue University, Montr\'{e}al, QC, H3A 2T8, Canada\\
\and
MilliLab, VTT Technical Research Centre of Finland, Tietotie 3, Espoo, Finland\\
\and
Moscow Institute of Physics and Technology, Dolgoprudny, Institutsky per., 9, 141700, Russia\\
\and
National University of Ireland, Department of Experimental Physics, Maynooth, Co. Kildare, Ireland\\
\and
Niels Bohr Institute, Blegdamsvej 17, Copenhagen, Denmark\\
\and
Observational Cosmology, Mail Stop 367-17, California Institute of Technology, Pasadena, CA, 91125, U.S.A.\\
\and
Optical Science Laboratory, University College London, Gower Street, London, U.K.\\
\and
SB-ITP-LPPC, EPFL, CH-1015, Lausanne, Switzerland\\
\and
SISSA, Astrophysics Sector, via Bonomea 265, 34136, Trieste, Italy\\
\and
SUPA, Institute for Astronomy, University of Edinburgh, Royal Observatory, Blackford Hill, Edinburgh EH9 3HJ, U.K.\\
\and
School of Physics and Astronomy, Cardiff University, Queens Buildings, The Parade, Cardiff, CF24 3AA, U.K.\\
\and
Space Research Institute (IKI), Russian Academy of Sciences, Profsoyuznaya Str, 84/32, Moscow, 117997, Russia\\
\and
Space Sciences Laboratory, University of California, Berkeley, California, U.S.A.\\
\and
Special Astrophysical Observatory, Russian Academy of Sciences, Nizhnij Arkhyz, Zelenchukskiy region, Karachai-Cherkessian Republic, 369167, Russia\\
\and
Stanford University, Dept of Physics, Varian Physics Bldg, 382 Via Pueblo Mall, Stanford, California, U.S.A.\\
\and
Sub-Department of Astrophysics, University of Oxford, Keble Road, Oxford OX1 3RH, U.K.\\
% ...
}

\title{\textit{Planck} 2013 results. XXIX. The \textit{Planck} catalogue of
Sunyaev--Zeldovich sources: \textit{Addendum} }
\authorrunning{Planck Collaboration}
\titlerunning{\textit{Addendum.} \textit{Planck} catalogue of Sunyaev--Zeldovich sources}


\abstract {We update the all-sky \Planck\ catalogue of 1227 clusters and
  cluster candidates (PSZ1) published in March 2013, derived from Sunyaev--Zeldovich (SZ) effect
  detections using the first 15.5 months of \Planck\ satellite
  observations. 
  \textit{addendum}, we deliver an updated version of the PSZ1
  catalogue, reporting the further confirmation of 86
  \Planck-discovered clusters.  In total, the PSZ1 now contains 947
  confirmed clusters, of which 214 were confirmed as newly discovered
  clusters through follow-up observations undertaken by the
  \Planck\ Collaboration. The updated PSZ1 contains redshifts for 913
  systems, of which 736 ($\sim 80.6$\%) are spectroscopic, and
  associated mass estimates derived from the $Y_z$ mass proxy.  We
  also provide a new SZ quality flag, derived from a novel artificial
  neural network classification of the SZ signal, for the remaining
  280 candidates. Based on this assessment, the purity of the updated
  PSZ1 catalogue is estimated to be 94\%. In this release, we provide
  the full updated catalogue and an additional readme file with
  further information on the \planck\ SZ detections. }
  
 \keywords{large-scale structure of Universe -- Galaxies: clusters: general --
   Catalogs}  
\maketitle


\section{Introduction}

\begin{figure*}[ht]
\begin{center}
\includegraphics{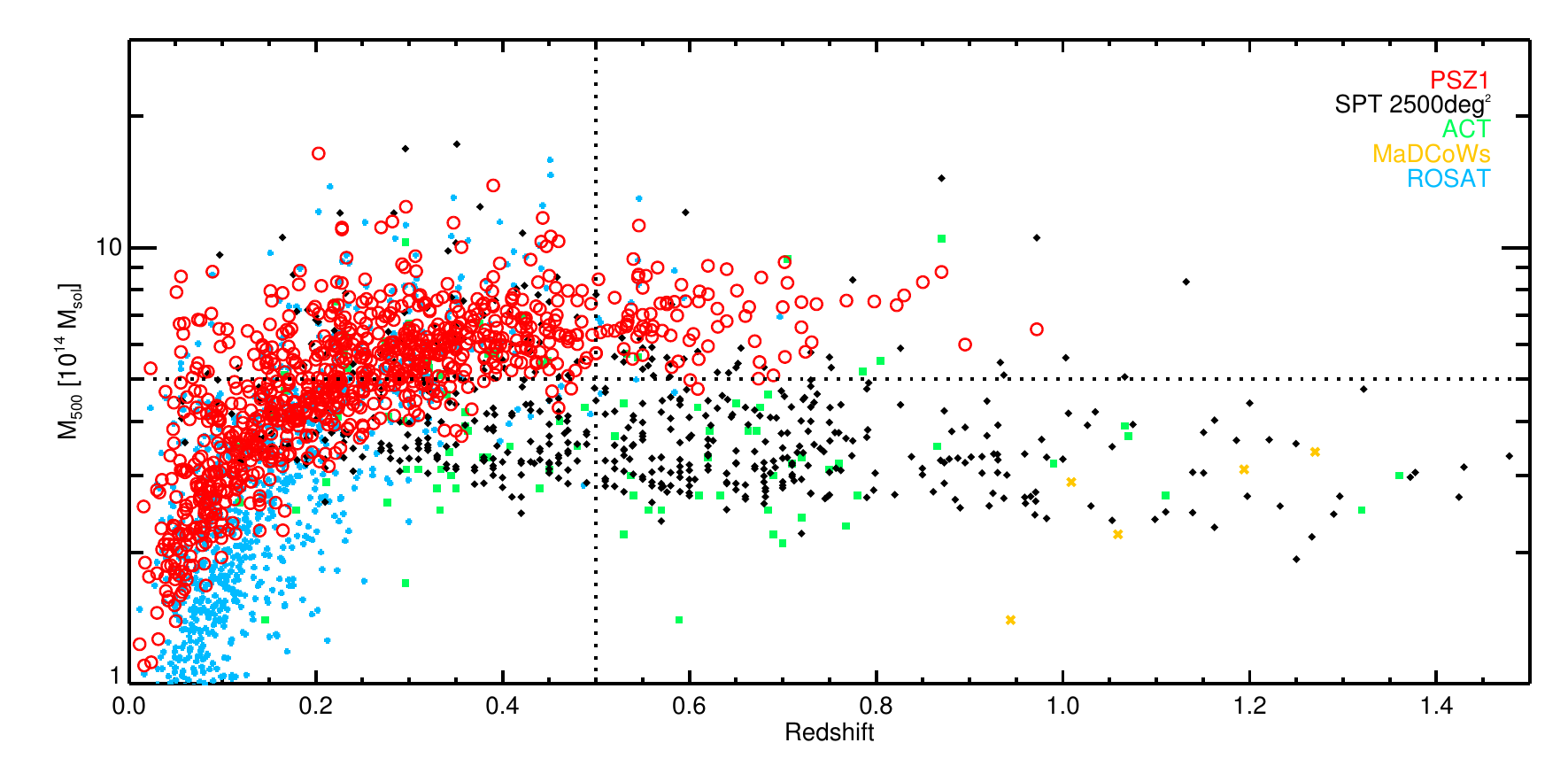}
\caption{{\footnotesize Distribution in the $M$--$z$ plane of the \planck\ SZ cluster catalogue
  (open red circles) \citep{planck2013-p05a} compared with those from SPT
  (black) \citep{rei13,ble14} and ACT (green)
  \citep{mar11,has13}, MaDCoWS (yellow) \citep{bro14}, and 
   NORAS and REFLEX from the MCXC meta-catalogue
  (blue) (\citet{pif11} and references therein). Some clusters may
  appear several times as distinct points due to differences in the
  mass estimate between surveys. The black dotted lines show the \planck\ mass limit
  for the medium-deep survey zone at 20\% completeness for a redshift limit of $z=0.5$. }}
\label{fig:mz}
\end{center}
\end{figure*}

It is only recently that cluster samples selected via their
Sunyaev--Zeldovich (SZ) signal have reached a significant sizes, e.g.,
the Early SZ (ESZ) catalogue from the
\Planck\ Satellite\footnote{\Planck\ (\url{http://www.esa.int/Planck})
  is a project of the European Space Agency (ESA) with instruments
  provided by two scientific consortia funded by ESA member states (in
  particular the lead countries France and Italy), with contributions
  from NASA (USA) and telescope reflectors provided by a collaboration
  between ESA and a scientific consortium led and funded by Denmark.}
\citep{planck2011-5.1a,planck2013-p05a}, and catalogues from the South Pole Telescope
\citep[ SPT, ][]{rei13,ble14} and the Atacama Cosmology Telescope
\citep[ACT,][]{mar11,has13}. These are now considered as 
new reference samples for cluster studies and associated cosmological analyses.

The present note describes updates to the construction and
properties of the \Planck\ catalogue of SZ sources PSZ1,
\citep[hereafter PXXIX2013, ][]{planck2013-p05a}, released in March 2013
as part of the first \Planck\ data delivery.  The PSZ1 catalogue
contains 1227 entries, including 683 so-called {\it previously-known}
clusters. This category corresponds to the association of \planck\ SZ
source detections with known clusters from the literature. The association is set to
the first identifier as defined in the hierarchy adopted by
PXXIX2013, namely: (i) identification with MCXC clusters
\citep{pif11}; (ii) identification with Abell and Zwicky objects; (iii) identification with clusters derived from SDSS-based
catalogues (primarily from \citet{wen12}); (iv) identification with clusters from SZ catalogues
\citep{has13,rei13}; (v) searches in the NED and SIMBAD
databases. Considerable added value,  including consolidated
redshift and mass estimates (Fig.~\ref{fig:mz}), has been obtained through
compilation of this ancillary information.

Since its delivery March 2013, we have continued to update the PSZ1 catalogue by
focusing on the confirmation of newly-discovered clusters in PSZ1. This process 
has first involved updating the redshifts of some previously-known clusters (Sect.~\ref{asso}). 
We have also made use of recent results from
dedicated follow-up observations conducted by the {\it Planck}
Collaboration with the RTT150 \citep{rtt150} and ENO telescopes 
(Planck Collaboration 2015, in prep.), which together have 
allowed us to observe and measure redshifts for $\sim 150$ PSZ1 sources (Sect. ~\ref{rtt}).  
We have also used published results from PanSTARRS \citep{liu14} and from the latest SPT catalogue \citep{ble14}, as described in Sects.~\ref{pans} and~\ref{spt}.   
For all clusters with measured redshifts, we have computed the 
estimated masses using the $Y_z$ mass proxy (\citealt{arn14} and PXXIX2013; Sec.~\ref{masses}). 
Finally, we have revisited the cluster candidate classification scheme, which in PXXIX2013 was organised into three classes ({\it class-}1, 2, 3) in order of decreasing reliability. 
As described in Sect. \ref{can}, we have now used the SZ spectral
energy distribution (SED) to refine the quality assessment of the
cluster candidates by adopting a new, novel quality flag derived from the Artificial
Neural Network analysis developed by \citet{agh14}.


\section{Redshift updates for \textit{previously-known} clusters}\label{asso}

In the external validation process performed in PXXIX2013, a total of
683 PSZ1 sources were associated with clusters published in X-ray,
optical, or SZ catalogues, or with clusters found in the NED or SIMBAD
databases. We refer to these as {\it previously-known clusters}.  Their
redshifts, when available, were compiled from the literature and
  a consolidated value was provided with the PSZ1 catalogue. 
  In the present update, we first re-examine the {\it
  previously-known} clusters of the PSZ1 catalogue.

The dedicated follow-up of \Planck\ PSZ1 clusters with RTT150
described in \citet{rtt150} provided updates to the redshifts of 19
{\it previously-known} clusters. The follow-up of
\Planck\ PSZ1 clusters with ENO telescopes further updated the
redshifts of five
  {\it previously-known} clusters. \\
We have updated the redshifts of ten
PSZ1 sources associated with SPT clusters provided in \citet{ble14}.
Finally, we have queried the NED and SIMBAD databases, and searched
in the cluster catalogues constructed from the SDSS data (namely
\citealt{wen12} and \citealt{roz14a}), for additional spectroscopic redshifts. When
these were available, we report them in the updated
version of the PSZ1 catalogue. The full process led us to change the  
redshifts of 34 {\it previously-known} PSZ1 clusters. We have
also changed the published photometric redshift estimate of one ACT cluster (ACT-CL
J0559-5249) to a spectroscopic redshift value.

In summary, 69
sources from the PSZ1 catalogue associated with {\it
  previously-known} clusters now have updated redshifts. Most of these consist of updates from photometric to spectroscopic values; however, eight redshifts were measured for the
first time for {\it previously-known} clusters.


\section{{\it Planck}-discovered clusters}

The PSZ1 catalogue contained 366 cluster candidates,
classified as {\it class-}1 to 3 in order of decreasing reliability,
and 178 {\it Planck}-discovered clusters confirmed mostly with
dedicated follow-up programmes undertaken by the {\it Planck}
Collaboration. Since the delivery of the PSZ1 catalogue in March 2013,
a number of additional confirmations, including results from the community, were
performed and redshifts were updated from photometric estimates to
spectroscopic values.

Combining the results from follow-up with the RTT150 \citet{rtt150}, ENO telescopes (Planck
collaboration 2015, in prep.), \citet{liu14}, \citet{roz14a}, and
\citet{ble14}, a total of 86 PSZ1 sources have been newly confirmed as {\it
  Planck-}discovered clusters with measured redshifts.

\subsection{From RTT150 results}\label{rtt}

As part of the \Planck\ Collaboration optical follow-up programme, candidates were observed with the Russian Turkish Telescope
\citep[RTT150\footnote{\url{http://hea.iki.rssi.ru/rtt150/en/index.php}.},][]{rtt150} within the Russian quota of observational time,
provided by Kazan Federal University and Space Research Institute
(IKI, Moscow). Direct images and spectroscopic redshift measurements
were obtained using T\"UB\.ITAK Faint Object Spectrograph and Camera
(TFOSC\footnote{\url{http://hea.iki.rssi.ru/rtt150/en/}\\ \url{index.php?page=tfosc}.}).
For the highest-redshift clusters, complementary spectroscopic
observations were performed with the BTA 6-m telescope of the SAO RAS
using the SCORPIO focal reducer and spectrometer \citep{afa05}.

These observations have confirmed and measured redshifts for a total of 24 new candidates. Eleven of these have spectroscopic redshifts. We have updated the PSZ1 catalogue by including these newly-measured redshifts. 

\subsection{From ENO telescopes}

Also as part of the \Planck\ Collaboration optical follow-up programme, candidates were observed
at European Northern Observatory (ENO\footnote{ENO:
  \url{http://www.iac.es/eno.php?lang=en}. }) telescopes, both in
imaging (at IAC80, INT and WHT) and spectroscopy (at NOT, GTC, INT and
TNG). The observations were obtained as part of proposals for the
Spanish CAT time, and an {\it International Time Programme (ITP)},
accepted by the International Scientific Committee of the Roque de los
Muchachos and Teide observatories. We summarise here the main results
of these observing programmes. Further details will be presented in a
companion article (\Planck\ Collaboration 2015, in prep.).

These observations have confirmed and provided new redshifts 
for a total of 26 candidates, that are reported in the updated
PSZ1 catalogue. These include the confirmation of 12 SZ sources as
newly-discovered clusters: two {\it class}-1, high reliability
candidates, five {\it class}-2, and five {\it class}-3 candidates.

\subsection{From PanSTARRS} \label{pans}

Based on the Panoramic Survey Telescope \& Rapid Response System
(PanSTARRS, \citealt{kai02}) data, \citet{liu14} have searched for
optical confirmation of the 237 \Planck\ SZ detections that overlap the
PanSTARRS footprint.

We only report here the redshifts for unambiguously confirmed clusters. Of these, 15 objects were  included in the RTT150 follow-up, for which the redshifts are published in
\citet{rtt150}, and three objects were included in the ESO follow-up described above. In these cases, we report the \Planck\ Collaboration follow-up redshift values in the updated PSZ1 catalogue. 
An additional two {\it Planck} clusters confirmed by
PanSTARRS have a counterpart in the \citet{roz14a} catalogue, with 
spectroscopic redshifts that we update in the PSZ1 catalogue.

A total of 40 {\it Planck-discovered} clusters are confirmed, for the
first time, by \citet{liu14} in the PanSTARRS survey.  All of these have
measured photometric redshifts that we have reported in the updated PSZ1 catalogue.

\begin{figure*}[!th]
\begin{center}
\includegraphics{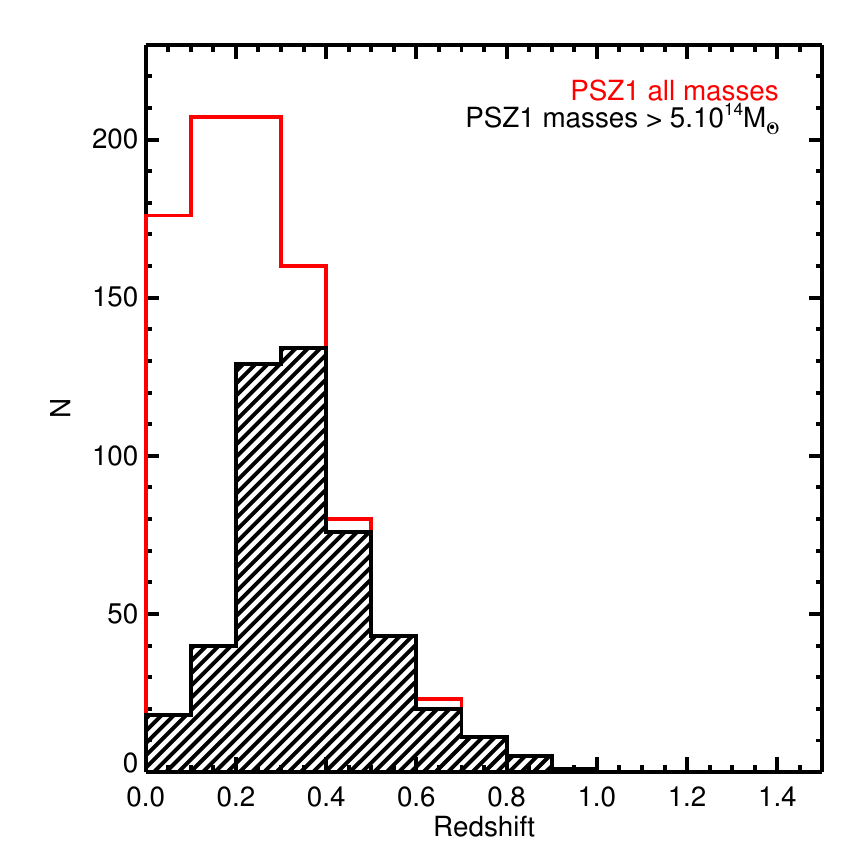}
\includegraphics{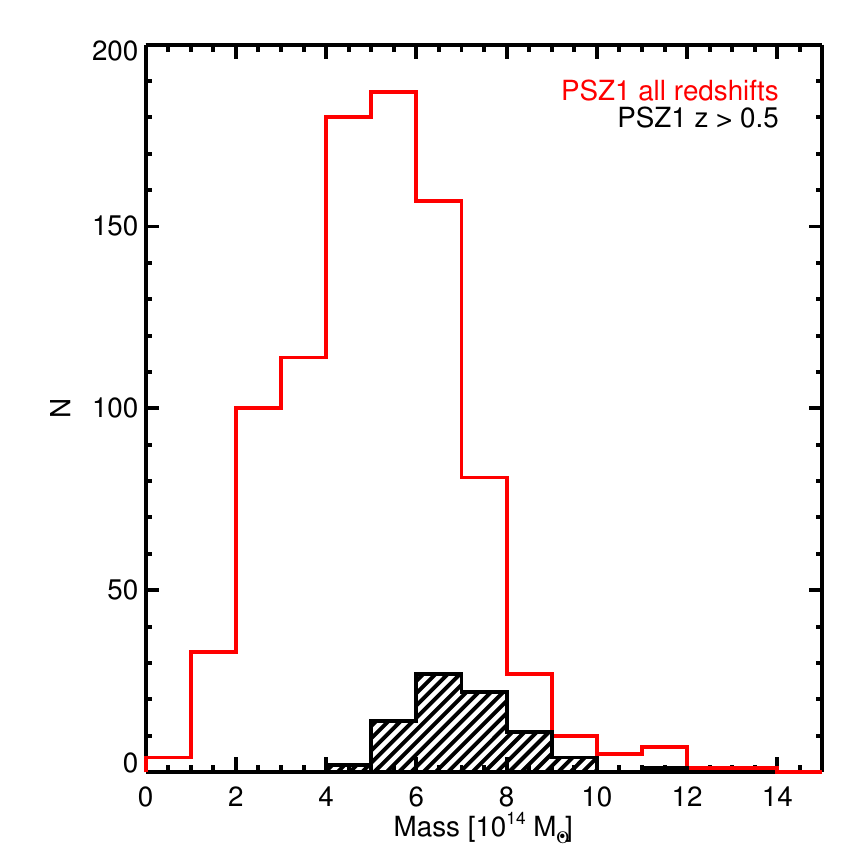}
\caption{{\footnotesize Distribution of redshifts (left panel) and masses (right
  panel) for the \planck\ SZ clusters. The black shaded area
  represents the population of clusters above redshift of 0.5 (in the
  right panel) and having a mass above $5 \times 10^{14} M_{\sun} $ (in the left
  panel).}}
\label{fig:z_hist}
\end{center}
\end{figure*}

\subsection{From SPT} \label{spt}

A new catalogue of SZ clusters detected with the South Pole Telescope
(SPT) cluster catalogue was published in \citet{ble14}. It provides an
ensemble of spectroscopic and photometric redshifts.  Four candidate {\it class-}1 and 2
clusters from the PSZ1 catalogue were confirmed
and have photometric redshifts in \citet{ble14}.  These are included in the updated PSZ1 catalogue.

\subsection{From SDSS-RedMapper catalogue} \label{spt}

Comparison with the SDSS-based catalogue from \citet{roz14a} provided
confirmation and new redshift values for five {\it Planck-discovered} clusters. This
includes confirmation of two {\it Planck} cluster candidates (one {\it class-}2
and one {\it class-}3 candidate). We use the spectroscopic redshift values 
available in the \citet{roz14a} in the updated PSZ1 catalogue.

\begin{figure}[h]
\begin{center}
\includegraphics[width=8.8cm]{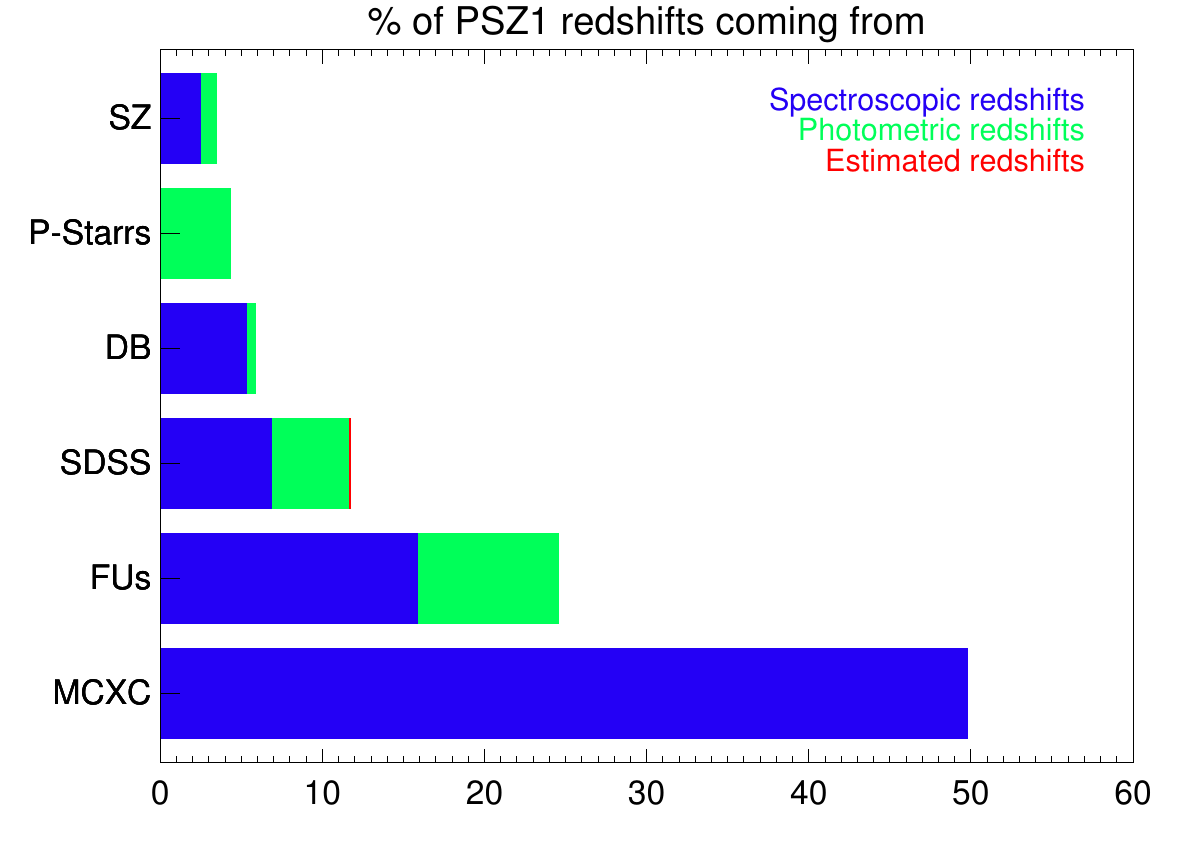}
\caption{{\footnotesize Percentage of origin and type (photometric, spectroscopic) of
  the redshifts reported in PSZ1. To date associations with MCXC
  clusters provide 49.8\% of the redshifts, all spectroscopic. Follow
  up observations by the \Planck\ collaboration (FUs) provide 24.6\% of the
  redshifts, of which 64.73\% are spectroscopic. Associations with clusters
  from SDSS-based catalogues result in 11.7\% of all redshifts, of which
  58.9\% are spectroscopic. Redshifts from the NED and SIMBAD databases
  represent 5.9\% of all redshifts, with 90.7\% of them spectroscopic. The confirmation
  from PanSTARRS data provides 4.4\% of the total number of redshifts, all of them
  photometric. Finally the association with SZ catalogues (SPT and
  ACT) represents 3.5\% of all redshifts, of which 71.9\% are spectroscopic.
}}
\label{fig:zdist}
\end{center}
\end{figure}


\section{Mass estimate} \label{masses}

The size--flux degeneracy discussed in, e.g., \citet{planck2011-5.1a}
and PXXIX2013 can be broken when $z$ is known, using the
\MYSZ\ relation between $\tv$ and $\YSZ$ see \citep{arn14}. The $\YSZ$
parameter, denoted $Y_z$, is derived from the intersection of the
\MYSZ\ relation and the size--flux degeneracy curve. It is the SZ mass
proxy $Y_z$ that is equivalent to the X-ray mass proxy $\YX$.

For all the \planck\ clusters with measured redshifts, $Y_z$ was
computed for our assumed cosmology, allowing us to derive an homogeneously-defined 
SZ mass proxy, denoted $M_{500}^{Y_z}$, based on X-ray
calibration of the scaling relations (see discussion in PXXIX2013). 
We show in Fig.~\ref{fig:z_hist}
(right panel, in red) the distribution of masses obtained from the SZ-based mass
proxy for all clusters with measured redshift. Note that since we use an X-ray calibration of the scaling relations, these masses are uncorrected for any bias due to the assumption of hydrostatic equilibrium in the X-ray mass analysis. The
shaded black area shows the distribution of masses for clusters with
redshifts higher than 0.5. They represent a total of 78 clusters.

\section{Cluster candidates}\label{can}

Since the delivery of the {\it Planck} catalogue and the confirmation in this \textit{addendum} 
of 86 candidates as new clusters, the updated
PSZ1 catalogue now contains 280 cluster candidates. In the original PSZ1, these latter were
classified as {\it class-}1 to 3 in order of decreasing reliability;
the reliability being defined empirically from the combination
of internal {\it Planck} quality assessment and ancillary information
(e.g., searches in RASS, WISE, SDSS data). The updated PSZ1
catalogue contains 24 high quality ({\it class-}1) SZ detections
whereas lower reliability {\it class}-2 and 3 candidates represent
130 and 126 SZ sources, respectively.

With the updated PSZ1 catalogue, we now provide a new objective quality
assessment of the SZ sources derived from an artificial neural-network
analysis. The construction, training and validation of the network is
based on the analysis of the Spectral Energy Distribution (SED) of the
SZ signal in the {\it Planck} channels. The implementation is discussed in detail by
\citet{agh14}.  The neural network was trained with an ensemble of three
samples: the confirmed clusters in the PSZ1 calatogue representing
good/high-quality SZ signal; the {\it Planck} Catalogue of Compact
Sources source, representing the IR and radio-source induced
detections; and random positions on the sky as examples of
noise-induced, very low reliability, detections.

\begin{figure}[h]
\begin{center}
\includegraphics[width=8.8cm]{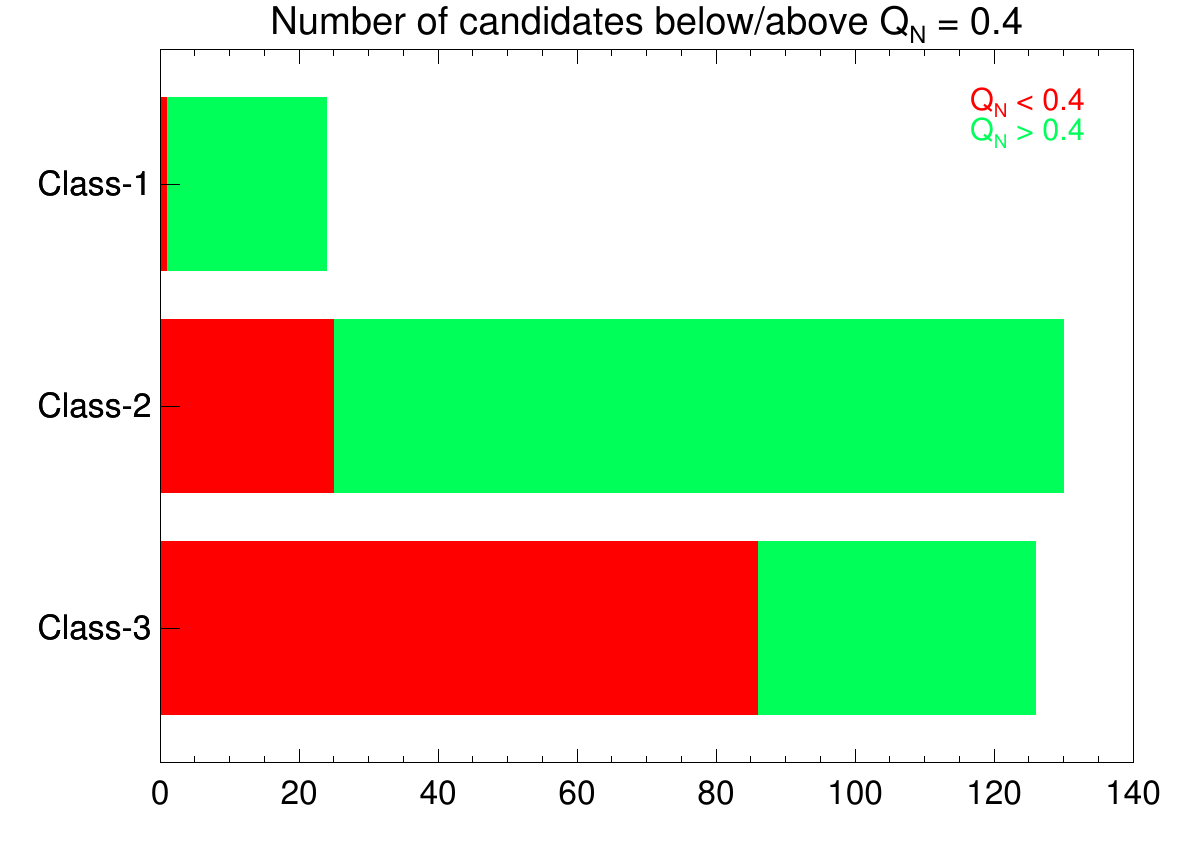}
\caption{{\footnotesize Number of {\it Planck} cluster-candidates below and above the
  neural-network quality flag threshold $Q_N=0.4$, denoting a high-quality SZ detection, for each reliability class.}}
\label{fig:zqual}
\end{center}
\end{figure}

In practice, we provide for each SZ source of the updated PSZ1 catalogue a
neural-network quality flag, $Q_N$, defined as in \citet{agh14}. This
flag separates the high quality SZ detections from the low quality
sources such that $Q_N< 0.4$ identifies low-reliability SZ sources
with a high degree of success.  Figure \ref{fig:zqual} summarises for
each class of {\it Planck} cluster candidate the number of sources
below and above the threshold velue of $Q_N=0.4$. The {\it class-}1 cluster
candidates all have $Q_N>0.4$ except for one source for which
$Q_N=0.39$. The fraction of `good' $Q_N>0.4$ SZ detections in the
{\it class-}2 category is about 80\%, while the fraction of $Q_N>0.4$ candidates 
drops to about 30\% for the {\it class-}3 cluster-candidates.

\section{Summary}

\begin{table*}[t]
\begingroup
\caption{{\footnotesize Summary of the updates of PSZ1v2 for each cluster or candidate type}}
\label{tab:summ}
\nointerlineskip
\vskip -3mm
\footnotesize
\setbox\tablebox=\vbox{
 \newdimen\digitwidth 
 \setbox0=\hbox{\rm 0} 
 \digitwidth=\wd0 
 \catcode`*=\active 
 \def*{\kern\digitwidth}
 \newdimen\signwidth 
 \setbox0=\hbox{+} 
 \signwidth=\wd0 
 \catcode`!=\active 
 \def!{\kern\signwidth}
\halign{$#$\hfil\tabskip 2em&
    \hfil$#$\hfil\tabskip=0.8em&
    $#$\hfil&
    \hfil$#$\tabskip=2em&
    \hfil$#$\hfil\tabskip=2em&
    \hfil$#$\hfil\tabskip=2em&
    $#$\hfil\tabskip=0.0em&
    \hfil$#$&
    \hfil$#$\tabskip=0pt\cr 
\noalign{\doubleline}
\hfil\hfil\hfil\hfil\hfil\hfil\hfil\hfil\hfil&\multispan3\hfil PSZ1 (2013)\hfil& \omit&\multispan3\hfil PSZ1v2 (2015)\hfil\cr
\noalign{\vskip -3pt}
 \omit\hfil\hfil&\multispan3\hrulefill&\omit&\multispan3\hrulefill\cr
\omit\hfil&\hfil Number\hfil&\multispan2\hfil redshift\hfil&\mathrm{UPDATES}&Number&\multispan2\hfil redshift\hfil\cr
 \omit\hfil\hfil&\omit&\multispan2\hrulefill&\omit&\omit&\multispan2\hrulefill\cr
\omit\hfil&\omit&type&number&\omit&\omit&type&number\cr
\noalign{\vskip 3pt\hrule\vskip 5pt}
\mathit{ `` Previously\; known'' } & \mathbf{683} & \omit&\omit&  \mathbf{0}  & \mathbf{683} &\omit&\omit\cr 
\omit                              & \omit & \mathrm{undef}&29& -8 & \omit& \mathrm{undef}& 21\cr 
\omit                              & \omit & \mathrm{estim}&5& -4 & \omit& \mathrm{estim}& 1\cr 
\omit                              & \omit & \mathrm{phot}&97& -43 & \omit& \mathrm{phot}& 54\cr 
\omit                              & \omit & \mathrm{spec}&552& +55 & \omit& \mathrm{spec}& 607\cr 
 \noalign{\vskip 6pt\hrule\vskip 5pt}
\mathit{ `` Planck\; discovered'' } & \mathbf{178} & \omit&\omit&  \mathbf{+86}  & \mathbf{264} &\omit&\omit\cr 
\omit                              & \omit & \mathrm{undef}&19& -6 & \omit& \mathrm{undef}& 13\cr 
\omit                              & \omit & \mathrm{phot}&72& +50 & \omit& \mathrm{phot}& 122\cr 
\omit                              & \omit & \mathrm{spec}&87& +42 & \omit& \mathrm{spec}& 129\cr 
 \noalign{\vskip 6pt\hrule\vskip 5pt}
 \mathit{Class-1 }                 & \mathbf{54} & \omit&\omit&  \mathbf{-30}  & \mathbf{24} &\omit&\omit\cr 
\omit                              & \omit & \mathrm{undef}&54& -30 & \omit& \mathrm{undef}& 24\cr 
\omit                              & \omit & \mathrm{phot}&\omit& \mathit{+22} & \omit& \mathrm{phot}& \omit\cr 
\omit                              & \omit & \mathrm{spec}&\omit& \mathit{+8} & \omit& \mathrm{spec}& \omit\cr 
\mathit{Class-2 }                  & \mathbf{170} & \omit&\omit&  \mathbf{-40}  & \mathbf{130} &\omit&\omit\cr 
\omit                              & \omit & \mathrm{undef}&170& -40 & \omit& \mathrm{undef}& 130\cr 
\omit                              & \omit & \mathrm{phot}&\omit& \mathit{+26} & \omit& \mathrm{phot}& \omit\cr 
\omit                              & \omit & \mathrm{spec}&\omit& \mathit{+14} & \omit& \mathrm{spec}& \omit\cr 
\mathit{Class-3 }                  & \mathbf{142} & \omit&\omit&  \mathbf{-16} & \mathbf{126} &\omit&\omit\cr 
\omit                              & \omit & \mathrm{undef}&142& -16 & \omit& \mathrm{undef}& 126\cr 
\omit                              & \omit & \mathrm{phot}&\omit& \mathit{+10} & \omit& \mathrm{phot}& \omit\cr 
\omit                              & \omit & \mathrm{spec}&\omit& \mathit{+6} & \omit& \mathrm{spec}& \omit\cr 
\noalign{\vskip 6pt\hrule\vskip 3pt}}}
\endPlancktablewide
\endgroup
\end{table*}

We have updated the \Planck\ catalogue of SZ-selected sources detected in
the first 15.5 months of observations. The catalogue contains 1227
detections and was validated using external X-ray and optical/NIR data,
alongside a multi-frequency follow-up programme for confirmation.

The updated PSZ1 catalogue now contains 947 confirmed clusters, including
264 brand-new clusters, of which 214 have been confirmed by the
\planck\ Collaboration's follow-up programme. The
remaining 280 cluster candidates have been divided into three classes
according to their reliability, i.e., the quality of evidence that
they are likely to be {\it bona fide} clusters. To date, high quality SZ
detections in PSZ1 represent 24 sources, all of which are classified as high-quality by our neural-network quality assessment procedure. Lower reliability, {\it class}-2
and 3 candidates represent 130 and 126 SZ sources respectively
(Table~\ref{tab:summ}). We find that $\sim 80$\% of the {\it class}-2
candidates are classified as high-quality by our neural-network quality assessment procedure, whereas only 35\% of the {\it class}-3 sources are considered as high-quality SZ
detections. Based on this assessement, the purity of the updated PSZ1
catalogue is $\sim 94\%$.

A total of 913 \planck\ clusters (i.e., 74.2\% of all SZ detections) now have measured redshifts, of which  736 are spectroscopic values (i.e., 80.6\% of all
redshifts).  The left-hand panel of Fig.~\ref{fig:z_hist} shows the distribution
in redshift of all clusters (red), and for the clusters with masses
above $5 \times 10^{14} M_{\sun}$ (shaded black). The median redshift of the
PSZ1 catalogue is about 0.23, and of order 35\% of the
\Planck\ clusters lie at redshifts above $z=0.3$.

The origins and types of redshifts are shown in
Fig.~\ref{fig:zdist}. Association with MCXC clusters
\citep{pif11} provides about $ 49.8\%$ of the redshifts, all of which are 
spectroscopic. Follow up observations undertaken by the \Planck\ Collaboration
provide $ 24.6\%$ of the redshifts, about two thirds of them
being spectroscopic. SDSS-based catalogues yield $ 11.7\%$
of the redshifts, of which more than half of which are spectroscopic. 
NED and SIMBAD database searches yield 5.9\% of the redshifts, the vast majority of which are  spectroscopic. PanSTARRS data provide $ 4.4\%$ of the redshifts, all of which are photometric. Finally, association with the SPT and ACT SZ catalogues represent $\sim 3.5\%$ of all 
redshifts, most of which are spectroscopic.

For the \planck\ clusters with measured redshifts, we have provided a 
homogeneously-defined mass estimated from the Compton $Y$
parameter. The $M$--$z$ distribution of the \Planck\ clusters is shown by open red circles 
in Fig.~\ref{fig:mz}, where it is compared with
other large cluster surveys. Note that the masses are not homogenised
and some clusters may appear several times due to differences in the
mass estimation methods between surveys.  We see that \planck\ cluster
distribution probes a unique region in the $M$--$z$ space occupied by
massive, $M\ge5\times 10^{14}\,\msol$, high-redshift ($z\ge 0.5$)
clusters.  The \Planck\ detections almost double the number of massive
clusters above redshift 0.5 with respect to other surveys.

\begin{acknowledgements}

The development of \Planck\ has been supported by: ESA; CNES and
CNRS/INSU-IN2P3-INP (France); ASI, CNR, and INAF (Italy); NASA and DoE
(USA); STFC and UKSA (UK); CSIC, MICINN, JA and RES (Spain); Tekes,
AoF and CSC (Finland); DLR and MPG (Germany); CSA (Canada); DTU Space
(Denmark); SER/SSO (Switzerland); RCN (Norway); SFI (Ireland);
FCT/MCTES (Portugal); and PRACE (EU). The authors thank
N. Schartel, ESA {\it XMM-Newton} project scientist, for granting the
DDT used for confirmation of SZ
\Planck\ candidates. The authors thank TUBITAK, IKI, KFU and AST for
support in using RTT150; in particular we thank KFU and IKI for
providing significant amount of their observing time at RTT150. We
also thank BTA 6-m telescope TAC for support of optical follow-up
project.  The authors acknowledge the use of the INT and WHT
telescopes operated on the island of La Palma by the Isaac Newton
Group of Telescopes at the Spanish Observatorio del Roque de los
Muchachos of the IAC; the NOT, operated on La Palma jointly by
Denmark, Finland, Iceland, Norway, and Sweden, at the Spanish
Observatorio del Roque de los Muchachos; the TNG, operated
on La Palma by the Fundacion Galileo Galilei of the INAF at the
Spanish Observatorio del Roque de los Muchachos; the GTC
telescope, operated on La Palma by the IAC at the Spanish Observatorio
del Roque de los Muchachos; and the IAC80 telescope
operated on the island of Tenerife by the IAC at the Spanish
Observatorio del Teide. Part of this research has been
carried out with telescope time awarded by the CCI International Time
Programme. The authors thank the TAC of the MPG/ESO-2.2m telescope for
support of optical follow-up with WFI under {\it Max Planck} time.
Observations were also conducted with ESO NTT at the La Silla Paranal
Observatory. This research has made use of SDSS-III data. Funding for
SDSS-III \url{http://www.sdss3.org/} has been provided by the Alfred
P. Sloan Foundation, the Participating Institutions, the National
Science Foundation, and DoE. SDSS-III is managed by the Astrophysical
Research Consortium for the Participating Institutions of the SDSS-III
Collaboration. \\ This research has made use of the following
databases: the NED and IRSA databases, operated by the Jet Propulsion
Laboratory, California Institute of Technology, under contract with
the NASA; SIMBAD, operated at CDS, Strasbourg, France; SZ cluster
database \url{http://szcluster-db.ias.u-psud.fr} and SZ repository
operated by IDOC operated by IAS under contract with CNES and CNRS.

\end{acknowledgements}

\bibliographystyle{aa}

\bibliography{catupdate,Planck_bib}

\appendix
\section{Description of the updated PSZ1 catalogue}

The updated \planck\ catalogue of SZ sources is available at PLA\footnote{
\url{http://www.sciops.esa.int/index.php?page=}\\
\url{Planck\_Legacy\_Archive\&project=planck}} and SZ cluster
database\footnote{\url{http://szcluster-db.ias.u-psud.fr}}. 

The updated PSZ1 gathers in a single Table all the entries of the
delivered catalogue based mainly on the \planck\ data and of the
external validation information, based on ancillary data (Appendices B
and C of \citet{planck2013-p05a} respectively). It also contains
additional entries. It is provided in a fits format, together with a
readme file.\\

The updated catalogue contains, when available, cluster external
identifications\footnote{The external identification corresponds to
  the first identifier as defined in the external validation hierarchy
  adopted in \citet{planck2013-p05a}.} and consolidated redshifts. We
added two new entries namely the redshift type and the bibliographic
reference. The three entries associated with the consolidated redshift
reported in the catalogue are thus:

\begin{itemize}
\item Type of redshift:  a string  providing the different
cases.
\begin{description}
   \item{\tt undef}: undefined
     \item{\tt estim}: estimated from galaxies
magnitudes
\item{\tt phot}: photometric redshift
\item{\tt spec}: spectroscopic
redshifts
\end{description}
\item Source for redshift: an integer value representing the origin of
  the redshifts.
\begin{description}
   \item{-1:} No redshift available
   \item{1:} MCXC updated compilation 
   \item{2}: Databases NED and SIMBAD-CDS
   \item{3:} SDSS cluster catalogue from \citet{wen12}
   \item{4:} SDSS cluster catalogue from \cite{sza11}
   \item{5:} SPT 
   \item{6:} ACT 
   \item{7:} Search in SDSS galaxy catalogue from
     \Planck\ Collab. from Fromenteau PhD 2010 and Fromenteau et
     al. (private comm.)
   \item{8:} SDSS catalogue from \citet{roz14a}
   \item{10:} Pan-STARRS1 Survey confirmation
   \item{20:} XMM-Newton confirmation from \Planck\ Collab. 
   \item{50:} ENO confirmation from \Planck\ Collab.
   \item {60:} WFI-imaging confirmation from \Planck\ Collab.
   \item{65:} NTT-spectroscopic confirmation from Planck Collab.
   \item{500:} RTT confirmation from \Planck\ Collab.
   \item{600:} NOT confirmation from \Planck\ Collab.
   \item{650:} GEMINI-spectroscopic confirmation from \Planck\ Collab.
\end{description}
\item Bibliographical references for the redshift
\end{itemize}

We also added a new entry describing further the quality of the SZ
detection. This is the flag $Q_N$ derived from the artificial neural
  network SED-based quality assesssment described in \cite{agh14}.


\raggedright
\end{document}